\documentclass[prl,showpacs,twocolumn,aps]{revtex4-1}

\usepackage{amssymb}
\usepackage{latexsym}
\usepackage{amsfonts}
\usepackage{epsfig}
\usepackage{psfrag}

\usepackage{epsfig,color}
\usepackage{psfrag}
\usepackage[utf8]{inputenc} 
\definecolor{darkgreen}{rgb}{0,0.4,0}

\newcommand{\bea}{\begin{eqnarray}}
\newcommand{\ea}{\end{eqnarray}}
\newcommand{\eea}{\end{eqnarray}}
\newcommand{\ord}{\,{\cal O}}

\newcommand{\tr}{\,{\rm Tr}}
\newcommand{\vc}[1]{\mathbf{#1}}

\usepackage{amsmath}

\usepackage{color}

\begin{document}

\title{Propagation of quantum fluctuations in quantum Ising model} 

\author{P. Navez$^{1,2}$, G. Tsironis$^{1,2}$, A. Zagoskin$^{2,3}$}

\affiliation{
$^1$ Crete Center for Quantum Complexity and Nanotechnology,
Department of Physics, University of Crete and FORTH, P. O. Box 2208, Heraklion 70113, Crete
\\
$^2$
National University of Science and Technology MISiS, Leninsky prosp. 4, 
119049 Moscow, Russia
\\
$^3$ Department of Physics, Loughborough University,
Loughborough LE11 3TU, United Kingdom
}

\date{\today}

\begin{abstract}
We investigate the entanglement dynamics between two distant qubits by analyzing correlations 
in the quantum Ising model. Starting from the spin system in a paramagnetic regime 
enforced by the external magnetic field $B$, we then switch on the ferromagnetic 
spin-spin coupling $J$. Using the large coordination number expansion, we consider 
two limiting switching regimes: (1) adiabatic, which monitors the evolution of the ground 
state through the quantum transition to an ordered state; and (2) instantaneous (quench) 
which monitors instead the propagation of quantum fluctuations and simulates the generation 
of long range correlations. In particular, we find that quantum fluctuations propagate with 
twice the group speed of excitations in the equilibrium state of the system.
\end{abstract}


\maketitle

{\it Introduction:}
The quantum Ising model has recently attracted additional attention 
as a standard generic model of quantum computers used to evaluate the behaviour of prototype devices\cite{google,exp2,exp3}. 
In particular, its study would considerably expand our understanding of both fundamental and practical 
limitations of adiabatic quantum computers and quantum annealers \cite{Zagos}, where the device is 
initiated in the strong transverse field, and then the spin-spin (qubit-qubit) coupling is gradually switched on. 
Entanglement between large number of spins on the intermediate stages of switching plays the key role in the system 
reaching its final ground state. In a real, open system the adiabatic evolution cannot take an arbitrary long time due 
to its eventual entanglement with the surroundings \cite{Sarandy}. The dynamics of entanglement is therefore 
crucially important for the operation of any quantum annealer.

While the final ground state of a quantum annealer is typically spin-glass like, 
some insight in this dynamics can be obtained in the simpler case of a  sweep through a 
symmetry-breaking quantum phase transition to the (anti)ferromagnetic order.   As the initial
quantum state is symmetric, all directions of symmetry breaking
are equally likely and seeded by quantum fluctuations.
Furthermore, the diverging response time at the critical point
indicates that the many-particle quantum system is driven
far away from equilibrium during the sweep. While nearby
points will most likely break the initial symmetry in the same
direction, two very distant points may spontaneously select
different directions of symmetry breaking \cite{KZ85}. As a result,
the spatial order parameter distribution after the quench will
be inhomogeneous and its spatial correlations are directly
determined by quantum correlations. 

The open questions in this context include: How is the order parameter
established and how fast does it spread? What is the role of these quantum fluctuations?  

In this Letter, we 
investigate the dynamics of the quantum Ising model both in the case of adiabatic 
 and instantaneous (quench) sweeping  with the help of the large coordination-number 
 expansion, which has been previously used  mainly in the context of a lattice Bose 
gas \cite{Brout,NS10,NQS14,QKNS14,KNQS14}. 
Since this model is only exactly solvable in 1D (one dimension) 
for nearest-neighbour couplings \cite{Pfeuty, HPK13},  for the Bethe lattice using DMRG 
\cite{NFG08}, or in 2D at the thermal equilibrium \cite{Liu}, 
we shall develop an alternative approach capable to handle 
nonequilibrium dynamics in higher dimension, with the possible applicability to different network connectivity.

After establishing the level of accuracy provided by our approach by 
comparizon to known exact solutions, we determine the dynamics of 
quantum fluctuations after a sudden quench to a fixed coupling value 
and  simulate the propagation of correlations inside the qubit system 
in the process of the ferromagnetic order formation.

{\it Quantum Ising model:}
We start from  the standard Hamiltonian
\cite{Pfeuty,NQS14}:  
\bea
\hat H=-\frac{J}{Z}\sum_{\mu\nu} T_{\mu\nu}
\hat{S}_{\mu}^z \hat{S}_{\nu}^z - B\sum_{\mu}\hat{S}_{\mu}^x 
\,.
\ea
where $\hat{S}_{\mu}^i$ ($i=x,y,z$)  are spin operators at sites $\mu,\nu$ interacting with each other with a coupling 
$J T_{\mu\nu}$ and placed in the transverse field $B$. The matrix $T_{\mu\nu}$ encodes the interactions 
in a $D$-dimensional hypercubic lattice of size $L$ with periodic boundary conditions  and is unity only for the nearest neighbours. 
In that case, the coordination number of the lattice is $Z=2D$.
This model displays a quantum transition from the paramagnetic phase 
(where the transverse magnetic field $B$ dominates) to the ferromagnetic 
or anti-ferromagnetic phase, which breaks the ${\mathbb Z}_2$ spin-flip 
symmetry. 

{\it Large coordination number expansion:} For $Z\gg1$, 
the model dynamics is described using the method  developed 
in \cite{NS10,QKNS14}. 
%We consider the hierarchy 
The time evolution of the density matrix $\hat\rho$ of the whole lattice is 
given by the von Neumann-Liouville equation 
$
i\hbar \partial_t\hat\rho 
=
\left[\hat H,\hat\rho\right] 
$.
This density matrix is usually too complex to be analyzed. Instead the 
set of reduced density matrices is introduced, 
$\hat\rho_S=\tr_{\not S}\hat \rho$,  which result from tracing out the Hilbert spaces of all 
sites except  a few: ${\cal S}=\{\mu_1,\mu_2, \dots, \mu_n\}$. If we keep only one site $\mu$, 
then the reduced density matrice is a linear operator $\hat \rho_\mu$ 
acting on the smaller Hilbert space 
of one lattice site $\mu$;  if we keep two sites $\mu,\nu$, then $\hat\rho_{\mu\nu}$ acts on the Hilbert space of two sites etc.
%These reduced objects simplify the determination of expected value of operators involving few sites,
%for example $\hat A_\mu$ has the expectation 
%$\langle \hat A_\mu \rangle= \tr_\mu (\hat A_\mu \hat \rho_\mu)$. 
%

The decomposition 
%Next we split up those matrices into correlated and uncorrelated parts. This decomposition 
%is the generalisation of  cumulant expansion used for probability distribution to density matrices. 
%For two sites, we write 
%
$\hat\rho_{\mu\nu}
=
\hat\rho_{\mu\nu}^{\rm corr}+\hat\rho_{\mu}\hat\rho_{\nu}
\,,
$
and 
$\hat\rho_{\mu\nu\lambda}=\hat\rho_{\mu\nu\lambda}^{\rm corr}+
\hat\rho_{\mu\nu}^{\rm corr}\hat\rho_{\lambda}+
\hat\rho_{\mu\lambda}^{\rm corr}\hat\rho_{\nu}+
\hat\rho_{\nu\lambda}^{\rm corr}\hat\rho_{\mu}+
\hat\rho_{\mu}\hat\rho_{\nu}\hat\rho_{\lambda}$ 
etc allows to derive an exact hierarchy of interlinked equations for these operators \cite{NS10,QKNS14}, the counterpart of the BBGKY chain. 

%an exact set of hierarchy equations has been 
%derived for these density matrices but in order to solve it, for instance,
%to obtain 
%$\hat\rho_{\mu}$, we would need the correlations 
%$\hat\rho^{\rm corr}_{\mu\kappa}$.
%
%Similarly, the equation for the two-point correlations 
%$\hat\rho^{\rm corr}_{\mu\nu}$ contains the three-point correlator 
%$\hat\rho^{\rm corr}_{\mu\nu\kappa}$ and so on.
Now we show if the initial state of the system is separable, then the 
correlations satisfy -- at least for a finite period of time -- 
the following hierarchy
$\hat \rho^{\rm corr}_{\cal S} \sim 1/Z^{|{\cal{S}}|-1}$, that is, the higher-order correlations are suppressed as an inverse power of the coordination number $Z$.  More explicitly, 
\bea \label{rho}
\label{hierarchy}
\hat\rho_{\mu} = \ord\left(Z^0\right)
,\,
\hat\rho^{\rm corr}_{\mu\nu} = \ord\left(1/Z\right)
,\,
\hat\rho^{\rm corr}_{\mu\nu\kappa} = \ord\left(1/Z^2\right)
,
\ea
and so on.
Using the spin representation, this statement can be rewritten as 
\begin{eqnarray}\label{dS}
S^{i}_{\mu}
=\langle \hat S^{i}_{\mu} \rangle &=& \ord \left(Z^0\right)
,\ \ 
M_{\mu \nu}^{ij}=\langle \delta \hat S^{i}_{\mu} \delta \hat S^{j}_{\nu} \rangle 
=\ord \left(1/Z\right)
,
\nonumber \\
\langle \delta \hat S^{i}_{\mu} \delta \hat S^{j}_{\nu} \delta \hat S^{k}_{\kappa} \rangle 
%\langle \delta \hat S^{i_1}_{\mu_1} \delta \hat S^{i_2}_{\mu_2} \dots \delta \hat S^{i_n}_{\mu_n} \rangle(t)
&=& \ord \left(1/Z^{2}\right), \ \ \dots  \quad \quad  i,j,k=x,y,z~,
\end{eqnarray}
where  $\delta \hat A_\mu = \hat A_\mu -\langle \hat A_\mu \rangle$. THis forms the basis of the $1/Z$-expansion.
 
{\it Mean-field approach:} In the leading order in $1/Z$ 
we neglect the correlations in order to obtain closed,  
time-dependent nonlinear mean field equations:
\begin{eqnarray}
\partial_t S_{\mu}^z
&=&\displaystyle
\frac{B}{2}
\left(S_{\mu}^- - S_{\mu}^+ \right)
\, ,
\\ 
i\partial_t S_{\mu}^\pm 
&=& 
\pm 2J\sum_{\nu}\displaystyle \frac{ T_{\mu\nu}}{Z}
S_{\mu}^\pm  S_{\nu}^z 
\mp B S_{\mu}^z
\,
\end{eqnarray}
where $S_{\mu}^\pm=S_\mu^x \pm i S_\mu^y$.
The lowest-order ground state is the mean field solution
that minimizes the mean field energy 
$E_0=-B\sum_{\mu} S^x_\mu - J \sum_{\mu\nu} T_{\mu\nu} S^z_\mu  S^z_\nu/Z$
with the reduced density matrix corresponding to a pure state 
$\hat\rho^{0}_\mu = |\psi\rangle_\mu \langle \psi|$ with
$|\psi\rangle_\mu= c_{\uparrow} |\uparrow \rangle_\mu + 
c_{\downarrow} |\downarrow \rangle_\mu$.
%
%The variational ground state energy $E_0$ per site reads 
%
%\bea
%\frac{E_0}{N}=
%-\frac{J}{4}(|c_{\uparrow}|^2 -|c_{\downarrow}|^2)^2 -
%\frac{B}{2}(c^*_{\uparrow}c_{\downarrow}+c_{\uparrow}c^*_{\downarrow})
%\,.
%\ea
%
We find here two regimes separated by 
a critical point at $J_c=B$.
For $J<B$, the magnetic field controls the orientation of the spin so that 
the state is paramagnetic with $S_{\mu}^{z(0)}=0$ and 
$S_{\mu}^{x(0)}=1/2$ (assuming $B>0$). 
For $J>B$, on the other hand, we get a non-vanishing ferromagnetic order 
parameter $S_{\mu}^{z(0)}=\pm\sqrt{1-B^2/J^2}/2$
(breaking the ${\mathbb Z}_2$ spin-flip symmetry). 

By introducing a linear perturbation around the steady state
$S^{i}_\mu (t)= S^{i(0)}_\mu +  S^{i(1)}_{\mathbf k} 
e^{i({\mathbf k}.{\mathbf x_\mu}-\omega_{\mathbf k}t)}$, where $\mathbf x_\mu$ is the 
site position,
we find the excitation modes with the following dispersion relation: 
\bea
\label{spectrum-Ising}
\omega_{\mathbf{k}}
=
\pm\sqrt{4J^2\left(S_{\nu}^{z(0)}\right)^2+B^2-2BJT_{\mathbf{k}}
S_{\nu}^{x(0)}}
\,.
\ea
where we define the Fourier components 
$
T_{\mu\nu}=\frac{1}{L^D}\sum_\mathbf{k} e^{i\mathbf{k}.(\mathbf{x}_\mu -\mathbf{x}_\nu)}
T_\mathbf{k}$ whose the expression for nearest neighbours is $T_\mathbf{k}=\sum_{i=1}^D \cos(k_i)/D$.
In both  paramagnetic and ferromagnetic regimes, the spectrum is gapped, but 
it becomes gapless at the transition. 
%$J_c=B$. 
In contrast to this result,  in 1D the critical point is at $J_c=2B$ \cite{Pfeuty}. Thus the mean field description 
valid for large $Z$ displays all essential qualitative features of phase and excitation spectrum 
but nevertheless appears to be a classical description of the on-site spin vector. Only the next order 
terms will reflect the quantum fluctuations that form the seeds for the ferromagnetic order. 

%For the paramagnetic phase, this simplifies into 
%
%\bea
%\label{para}
%\omega_{\mathbf{k}}
%=
%\pm\sqrt{B^2-BJT_{\mathbf{k}}}
%\,.
%\ea
%
%In contrast to the Heisenberg model, some imaginary frequencies exist
%for $J>B$ and mark the transition to the ferromagnetic regime. 
%In the ferromagnetic case, the spectrum~(\ref{spectrum-Ising}) becomes 
%
%\bea
%\label{ferro}
%\omega_{\mathbf{k}}
%= 
%\pm\sqrt{J^2-B^2T_\vc{k}}
%\,.
%\ea
%
%Again, we see that  some of the frequencies acquire an imaginary part if
%we cross the critical point at $J=B$. 
%In contrast in 1D, the critical point is at $J=2B$ \cite{Pfeuty}.

{\it Adiabatic switching:} We start from the initial condition  $S_\mu^x=\frac{1}{2}$, 
$M_{\mu\nu}^{ij}=0$ which is an eigenstate of an Hamiltonian in absence of ferromagnetism at $J=0$. 
Then we switch $J(t)$ with some particular time dependence. With such uniform initial conditions, all the spins behave in the 
same way and the translational invariance imposes correlations 
depending only on the relative distance so that we define
$y(t)=2J(t) S_\mu^i(t)/B$ and 
$M_{\mu\nu}^{ij}=\frac{1}{L^D}\sum_\mathbf{k} e^{i\mathbf{k}.(\mathbf{x}_\mu -\mathbf{x}_\nu)}
M^{ij}_\mathbf{k}$.
The Fourier transform defines also the unphysical uniform correlations,
so that the physical correlations are
${\tilde{M}}^{ij}_\mathbf{k}=
M^{ij}_\mathbf{k}-\frac{1}{L^D}\sum_\mathbf{k'}M^{ij}_\mathbf{k'}$. Using the symmetry 
$M^{ij}_\mathbf{k}=M^{ji}_\mathbf{k}$, the non trivial equations of motion are derived 
from the next order equations of the hierarchy\cite{NS10,QKNS14}:
\begin{eqnarray}
\partial_t \left(\frac{y}{J}\right) &=&
\frac{4J}{B L^D}\sum_{\mathbf{k}} 
T_{\mathbf{k}}M^{yz}_{\mathbf{k}}~,
\\
\partial_t M_{\mathbf{k}}^{zz}&=&
-2BM_{\mathbf{k}}^{yz}~,
\\ \label{fxy}
\partial_t M_{\mathbf{k}}^{yz}&=&
-B M_{\mathbf{k}}^{yy} + B M_{\mathbf{k}}^{zz}
-BT_{\mathbf{k}}(\tilde{M}^{zz}_{\mathbf{k}}+\frac{1}{4})y~,
\\
\partial_t M_{\mathbf{k}}^{yy}&=&
2B M_{\mathbf{k}}^{yz} 
-2B T_{\mathbf{k}}\tilde{M}^{yz}_{\mathbf{k}}y~.
\end{eqnarray}

The ground state solution and quantum transition point are determined by an adiabatic switching with
the profile  $J(t)=J_c \exp(\epsilon t)$ in the interval $t=]-\infty, 0]$
%In order to follow the ground state adiabatically from the deep paramagnetic regime to the quantum transition 
%to the ferromagnetic phase, we 
with $\epsilon$ infinitesimally small. Noticing that 
$\partial_t =\epsilon J\partial_J$, the time parameter is eliminated from the dynamical equations.
%In the adiabatic limit ($\epsilon \rightarrow 0$), 
The introduction of the scaling $M^{yz}_{\mathbf{k}} \rightarrow \epsilon$ and the unity scale  
for all other dynamical variables together with the elimination of 
$M^{yz}_{\mathbf{k}}$ using (\ref{fxy}) result in the following $\epsilon$-independent equations:
%after taking the limit:  
\begin{eqnarray}\label{y}
\partial_J \left(\frac{y}{J}\right) &=&
-\frac{2J}{B^2 L^D}\sum_{\mathbf{k}} 
T_{\mathbf{k}}
\partial_J  M_{\mathbf{k}}^{zz}~,
\\
M_{\mathbf{k}}^{zz}&=&
M_{\mathbf{k}}^{yy} + T_{\mathbf{k}}(\tilde{M}^{zz}_{\mathbf{k}}+\frac{1}{4})y~,
\\
\partial_J  M_{\mathbf{k}}^{yy}&=&
-\partial_J M_{\mathbf{k}}^{zz} 
- T_{\mathbf{k}}y\partial_J \tilde{M}^{zz}_{\mathbf{k}}~.
\end{eqnarray}
This independence ensures that the spin system remains in its ground state 
by tuning  $J$, as long as this state is not degenerate. Only at the transition point 
the excitation spectrum becomes gapless and the smooth evolution is not guaranteed.
The elimination of $M^{yy}_{\mathbf{k}}$ from the two last equations leads to
a first order 
%\begin{eqnarray}
%(1-y T_{\mathbf{k}})\partial_J \tilde{f}_{\mathbf{k}}^{zz}
%-\frac{1}{2}T_{\mathbf{k}}(\partial_J y)\tilde{f}_{\mathbf{k}}^{zz}
%=\frac{1}{8}T_{\mathbf{k}}\partial_J y - \partial_J f^{zz}_{0}
%\end{eqnarray}
%where $f^{zz}_{0}=
%L^{-D}\sum_\mathbf{k'}f^{zz}_\mathbf{k'}$. This first order 
equation for $M^{zz}_\mathbf{k}$ that is solved using  the variable 
change $y(J)$.  We obtain:
\begin{eqnarray}\label{fk}
\tilde{M}_{\mathbf{k}}^{zz}(y)=
\frac{1}{\sqrt{1-yT_{\mathbf{k}}}}\left[
\frac{1}{4}-\int_0^{y}\!\!\!\!\!dy'\frac{\partial_{y'} f_{0}(y')}{\sqrt{1-y'T_{\mathbf{k}}}}
\right]-\frac{1}{4}
\end{eqnarray}
where $f_{0}=
L^{-D}\sum_\mathbf{k'}M^{zz}_\mathbf{k'}$. 
The application of the condition $L^{-D}\sum_\mathbf{k}\tilde{M}^{zz}_\mathbf{k}=0$ to this last equation provides 
a one-dimensional closed integral equation for $f_{0}(y)$:
\begin{eqnarray}\label{f0}
\frac{1}{L^{D}}\sum_\mathbf{k}
\frac{1}{\sqrt{1-yT_{\mathbf{k}}}}\left[
1-\int_0^{y}\!\!\!\!\! dy' \frac{4\partial_{y'} f_{0}(y')}{\sqrt{1-y'T_{\mathbf{k}}}}
\right]=1~.
\end{eqnarray}
Together with  Eq.(\ref{y}), they provide the spin solution with a low cost in terms 
of computation time
%for $S^x_\mu(J)=B y(J)/2J$
through integration: 
\begin{eqnarray}\label{S}
S^x_\mu(y)=\left[\frac{1}{4} - \int_0^y \!\!\!\!\!dy' y' 
\frac{1}{L^{D}}\sum_\mathbf{k} T_\mathbf{k} \partial_{y'} {\tilde M}^{zz}_\mathbf{k}(y')\right]^{1/2}.
\end{eqnarray}
The singularity appearing  at $y(J)=1$ corresponds to the quantum phase transition at $({S^x}_\mu)_c=(B/2J)_c$ 
up to the 
next order in $1/Z$. In that case 
the correlations become singular at $\mathbf{k}=0$. For nearest neighbour interactions, 
we obtain the long wavelength scaling $\tilde{M}_{\mathbf{k}}^{zz}(y=1)\rightarrow 1/|\mathbf{k}|$. 

The results are plotted in Fig.1 for the large system size
($L \rightarrow \infty$) and go beyond the mean-field results.
For $D=1$, there exist an analytical solution, and the next-order correction already converges quite well to 
the exact value, with the critical point given by:
%
%The spin is 
%determined by solving Eq.(\ref{f0}) from which we deduce successively 
%(\ref{fk}) and (\ref{S}).
%\begin{eqnarray}
%f_{0}(y=1)=\frac{1}{4}-\frac{1}{2\sqrt{2}\pi}(1+\frac{1}{2\sqrt{2}}\ln[3+2\sqrt{2}])
%\end{eqnarray}
%and the critical spin:
%
\begin{eqnarray}
%{S_\mu^x}^2=\frac{1}{2}-\frac{1}{2\sqrt{2}}[1-\frac{1}{2\pi}(-1+\frac{1}{2\sqrt{2}}\ln[3+2\sqrt{2}])]
({S^x_\mu})_c=\sqrt{\frac{\sqrt{2}}{\pi}-\frac{1}{4}}\simeq 0.44~.
\end{eqnarray}
This value is closer to the exact one $(S^x_\mu)_c=1/\pi=0.32$ in  \cite{Pfeuty} and 
shofts the critical coupling in the right direction, to $(J/B)_c=1.12$, which is nevertheless still below
 the exact value of $(J/B)_c=2$. 
For $D=2$, the critical value  $(J/B)_c=1.075$  approaches better the exact result $(J/B)_c=1.314$ obtained through the quantum 
Monte-Carlo method \cite{Liu}. For $D=3$ the obtained value $(J/B)_c=1.042$ is even closer to the mean field curve. It is reasonable to  
 expect it to better fit the exact results for larger coordination number.
 
We note that the transition takes place once the spin value reaches the mean field value in the ferromagnetic regime. 
The spin reduction is a signature of the global entanglement of a given site with all the others which can be estimated through 
the relation $\eta_\mu=2\left[1-\tr(\hat \rho_\mu^2)\right]=
1-4{S^x_\mu}^2$. It shows that the entanglement increases close to the transition point but decreases with the dimensionality of the system.
The Fig.2 displays the growing extension of the range of quantum fluctuations  as we approach the transition at which 
all sites becomes correlated. 

Generally, all the results obtained by adding the two-sites correlations 
improve over the mean-field approach  and reproduce correctly the physical properties of the ground state, especially in higher dimensions.

\begin{figure}[t]
\begin{center}
\includegraphics[width=8.5cm]{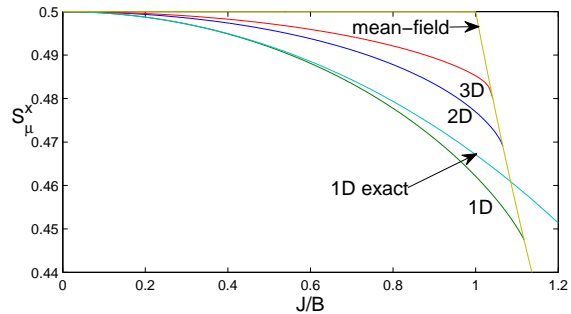}
% N=1000
\end{center}
\caption{Plot of the spin $S^x_\mu$ vs. $J/B$ in $D=1,2$ and $3$. The mean-field  
and 1D exact results are plotted for comparison.}
\end{figure}

\begin{figure}[t]
\begin{center}
\includegraphics[width=8.5cm]{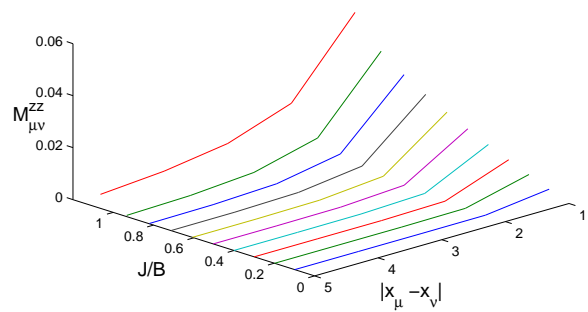}
\end{center}
\caption{
Plots of the quantum fluctuations correlator $M^{zz}_{\mu\nu}$ vs. $J/B$ and intersite distances $|{\bf x}_\mu - {\bf x}_\nu|$ 
in 2D ($L=40$).}
\end{figure}

%To go to the continuous case, we use the formulae

%\begin{eqnarray}
%DOS(E)=
%\sum_{\mathbf{k}} 
% \frac{1}{L^D}\delta(JT_{\mathbf{k}} -E)= \int_ 0^\infty dt\,I_0^D\left(\frac{Jt}{D}\right)e^{-Et}
%\end{eqnarray}

%{\color{red}

%The energy can be written:
%\begin{eqnarray}
%{\langle \hat H \rangle}
%=J L^D(1/4 -3{S_\mu^x}^2-2f_{0})/y
%\end{eqnarray}

%{\it Entanglement calculation:}
%The entanglement of the set $S$ with the rest of the system is 
%\begin{eqnarray}
%\eta_S=
%\frac{2^{|S|}}{2^{|S|}-1}\left[1-\tr(\rho_S^2)\right]
%\end{eqnarray}

%{\bf Case $S=\{\mu\}$:}

%The one-site reduced density matrix is given by $\rho_\mu=\frac{\mathbb{I}}{2}+\mathbf{\hat S}_\mu.\mathbf{S}_\mu$.
%The entanglement is in this case is: $\eta_\mu=1-4{S^x}^2$. %
%The maximum possible entanglement is in 1D and obtained at the transition. 

%{\bf Case $S=\{\mu,\nu\}$}

%The two-site reduced density matrix is given by:
%\begin{eqnarray}
%\hat \rho_{\mu\nu}=\hat \rho_\mu \hat \rho_\nu + M_{\mu\nu}^{yy}\mathbf{\hat S}_\mu^y \mathbf{\hat S}_\nu^y 
%+ M_{\mu\nu}^{zz}\mathbf{\hat S}_\mu^z \mathbf{\hat S}_\nu^z 
%\end{eqnarray}
%We obtain:
%\begin{eqnarray}
%\eta_{\mu \nu}=
%\frac{4}{3}\left[1-2{S^x}^2-1/4 -4({M_{\mu\nu}^{yy}}^2 +{M_{\mu\nu}^{zz}}^2+{S^x}^4))\right]
%\end{eqnarray}

%We should take into account the Cauchy-Schwartz inequality:
%\begin{eqnarray}
%CS= |\sqrt{{M_{\mu\nu}^{yy}}^2+ { M_{\mu\nu}^{zz}}^2}+{S^x}^2|\leq 1/4.
%\end{eqnarray}
%which unfortunately is violated for nearest neighbours. It shows the quality of the approximation.

%}

{\it Quench dynamics:}
Now we use our approach to describe the dynamics of quenching.   
We start from $J=0$ and we instantanously switch it to a final value $J$.
On the time scale much shorter than the decoherence time,  
the evolution of the spin system can be considered as entirely determinist when starting from the zero temperature ground state. 
Therefore any  fluctuations resulting from the quench are purely quantum over this time interval. 

In the paramagnetic regime ($J< J_c$), the transverse 
spin evolves to reach a steady value lower than the 
corresponding one for the ground state but still close to $1/2$.
We observe a wave pulse-like generation  of the correlations that propagate just after the quench. 
The  propagation speed is constant and estimated as $twice$ 
the group velocity Eq.(\ref{spectrum-Ising}) of the excitation in the system,
$c={\rm Max}\left(2\partial \omega_\vc{k}/\partial k_x \right)=2 (J/Z)[{1-J(Z-2)/(BZ)}]^{-1/2}$.
The factor two corresponds to spontaneous virtual excitations always created in pairs and 
follows from the physical reality that at least two kinks are needed for a spin domain formation.  
%0.8/(2\sqrt(0.6)=0.516
In the case of $J/B=0.8$ in Fig.3, we obtain the value  $c\simeq 0.5 B$.  
In a system with periodic boundary condition, the signal is reflected back as an echo that affects the transverse 
spin at $Bt=150$ with a small oscillating burst. The amplitude of the 
pulse is modulated by oscillations 
of frequency estimated as
$2B\sqrt{1-J/B}/2\pi$. 
%$2B\sqrt{1-2JS_\mu^x/B}/2\pi$. 
As a result of random waves, the correlations remain short-ranged and the total magnetization fluctuations  
$M^{zz}={\tilde M}_{\vc{k}=0}^{zz}$ along the $z$ axis 
cannot develop but instead stay confined to a small value as can be seen in the second graph of 
Fig.3. 

The visible wavefront can be exploited to test the quantum character of the interaction 
between the spin sites. It is an essentially quantum prediction, which goes beyond 
the mean-field approach.

In the case of a quench beyond the critical value, $J >J_c$, 
the sweep is done in the ferromagnetic region and quantum correlations develop over a long range.
The frequency of propagation becomes imaginary and leads to an exponential increase with a rate given by 
$\gamma=2\omega_{\vc{k}=0}/i=2B\sqrt{2JS_\mu^x /B-1}\sim (J-J_c)^{1/2}$. 
The second graph of Fig.4 shows indeed a growing $M^{zz}=M^{zz}_{max} \gamma t$ 
for $J/B=1.5$, where 
$M^{zz}_{max}=L^D/4$ corresponds to the maximum possible value of the fluctuations correlator, while 
the average spin stabilizes to a much lower steady state value.
Using the long-wavelength approximation  $2 \omega_{\vc{k}=0}=i\sqrt{\gamma^2 - v^2 {\mathbf k}^2}$
where $v=2JS_\mu^x B$, the saddle point method is used to estimate the scaling of
the onset of correlations \cite{NS10}:
\begin{eqnarray}
{M}^{zz}_{\mu\nu}\sim
\exp(\gamma \sqrt{t^2 -|{\mathbf x}_\mu -{\mathbf x}_\nu|^2/v^2})~.
\end{eqnarray}
This dominant term displays a finite propagation speed  $v$ for the onset of correlations shown
in Fig.4. It tells about the pre-stage dynamics towards a steady ferromagnetic state.
Nevertheless, this analysis is restricted to a short time scale within the region of 
the instability of the growing modes and thus cannot be extended to study 
the evolution towards the thermodynamic equilibrium. 

\begin{figure}[t]
\begin{center}
\includegraphics[width=4cm]{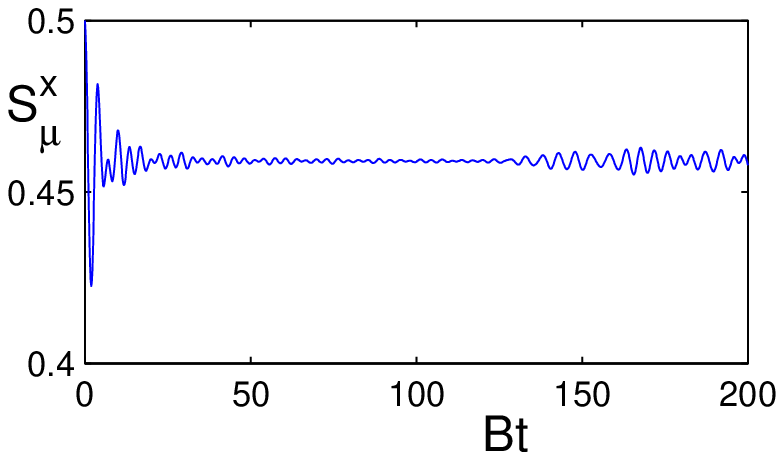}
\includegraphics[width=4cm]{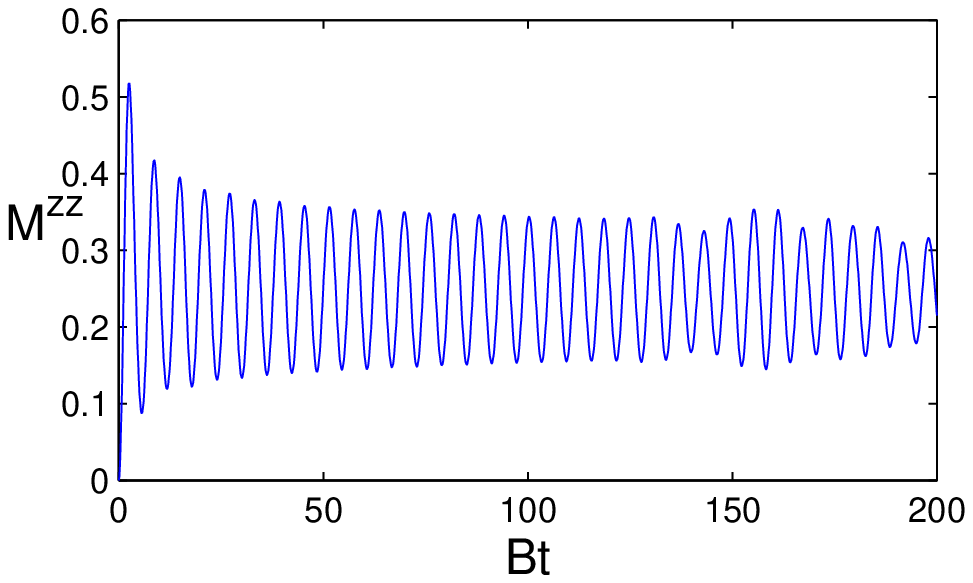}
\includegraphics[width=8.5cm]{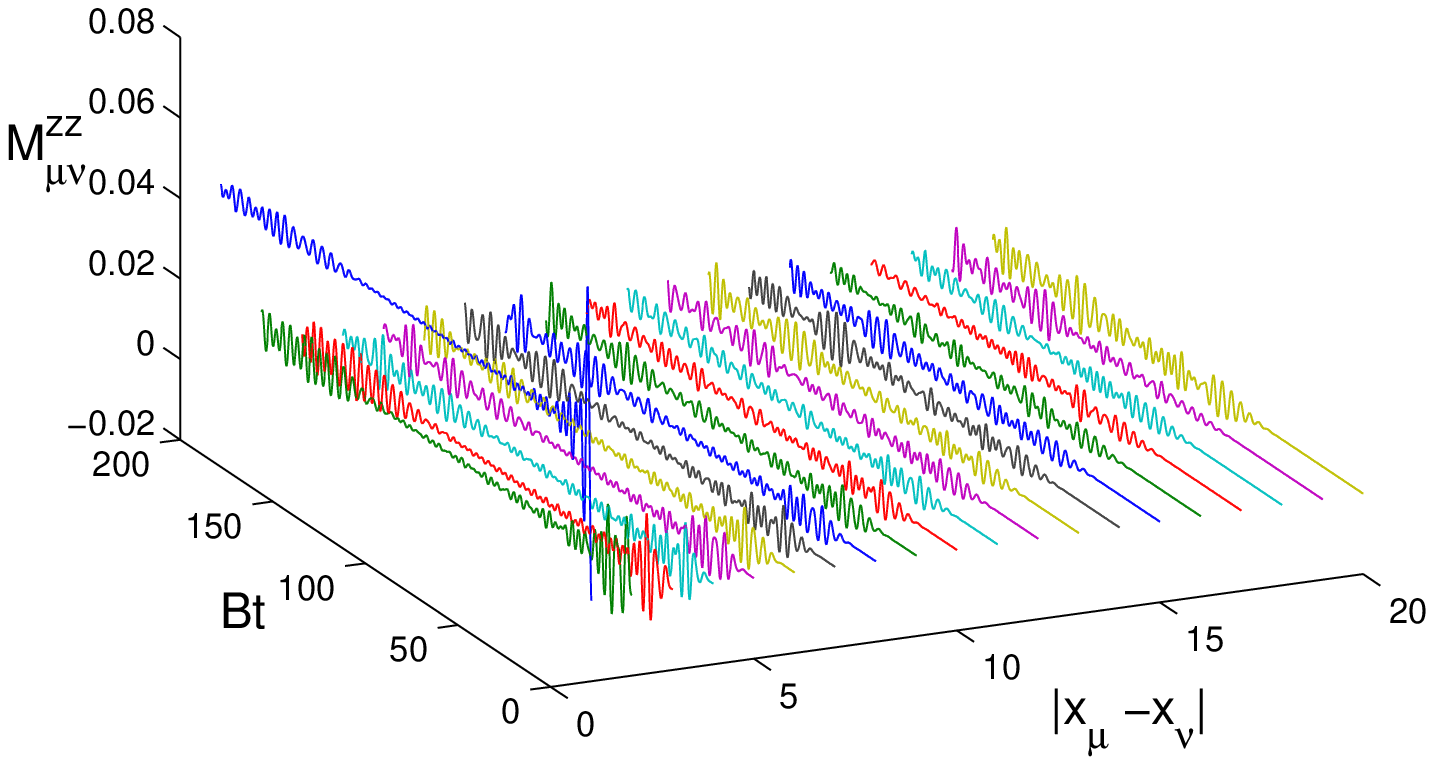}
% N=1000
\end{center}
\caption{
Quench dynamics: time evolution of spin $S^x_\mu$, the fluctuation correlator  $M^{zz}_{\bf k=0}$ and local fluctuation correlators $M_{\mu\nu}^{zz}$ along the principal axis for $J/B=0.8$ in 2D ($L=40$).}
\label{q08}
\end{figure}

\begin{figure}[t]
\begin{center} 
\includegraphics[width=4cm]{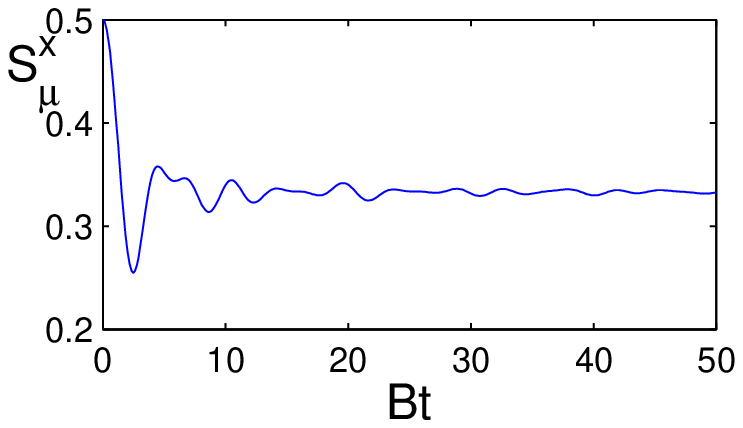}
\includegraphics[width=4cm]{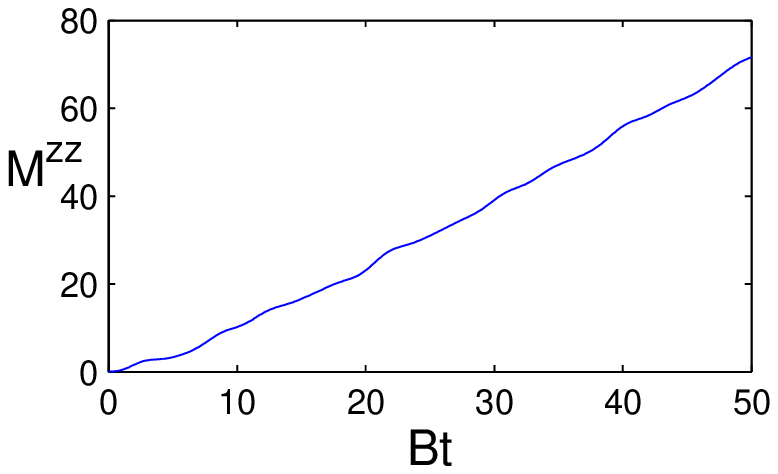}
\includegraphics[width=8.5cm]{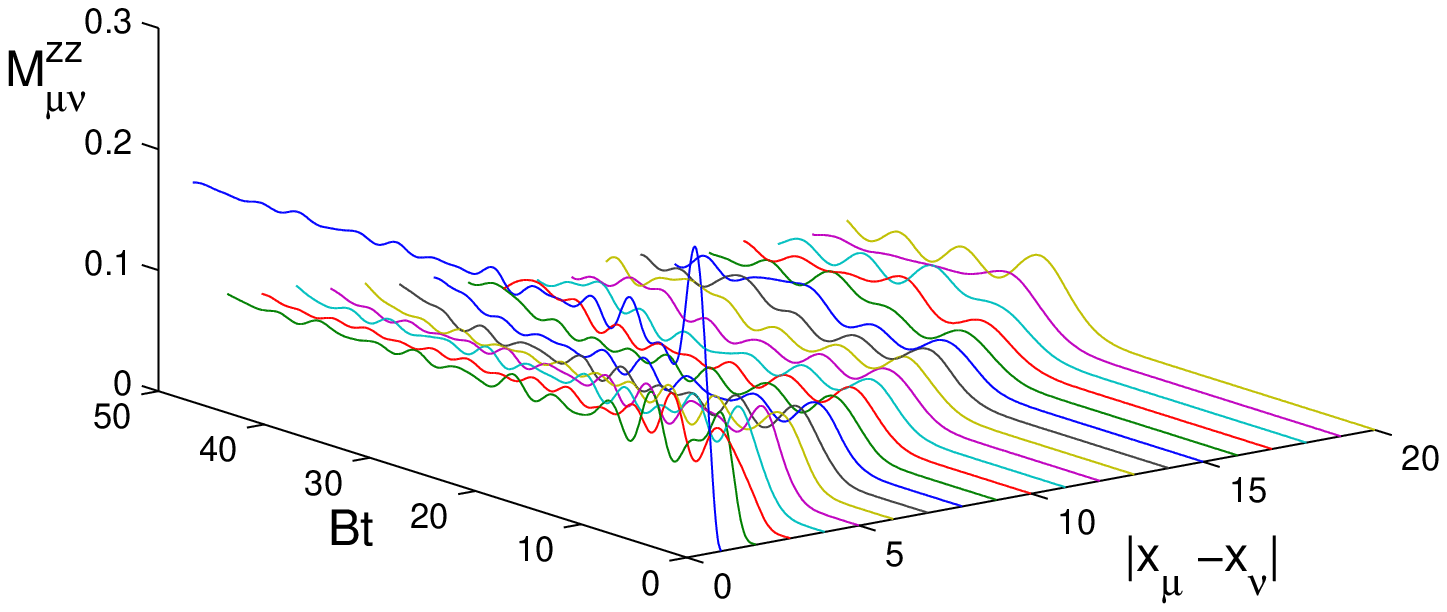}
% N=1000
\end{center}
\caption{
%Plots of time evolution of spin $S^x_\mu$, the generation of correlations $M^{zz}_{\bf k=0}$ 
Same as in Fig.3 but for $J/B=1.5$.}
\label{q15}
\end{figure}

{\it In summary:} We have used the large coordination number expansion developed in 
\cite{NS10,QKNS14}
in order to describe the nonequilibrium dynamics of the quantum Ising 
model. Since this technique is based
on the evolution equations of reduced density matrices, it can be used 
efficently for numerical simulations.
The lowest order in $1/Z$ reproduces the mean-field classical
approach; the higher orders describe quantum
correlations. This method has been applied to calculate
the creation and amplification of quantum correlations in
a quenched paramagnetic-ferromagnetic phase transition. We
find that the off-site long-range order spreads with
a constant velocity exceeding the excitation speed in the system in equilibrium. 
An experimental observation of this effect could provide arguments in favour of 
quantum dynamics of an artificial quantum structure such as a quantum annealer.
 
%and obeys universal scaling laws. 

\begin{acknowledgments}
We acknowledge partial supports of the European Union's Seventh
Framework Programme (FP7-REGPOT-2012-2013-1) under grant agreement number 316165, 
by the EPSRC grant EP/M006581/1 and by
the Ministry of
Education and Science of the Russian Federation in the 
framework of Increase Competitiveness Program of NUST MISiS No. K2-2014-015 and 
No. K2-2015-007.
Helpful discussions with J. Betouras, F. Queisser and K. Krutitsky are gratefully acknowledged.
\end{acknowledgments}

%----------------------------------------

\end{document}